\documentclass[aps,prl,twocolumn,notitlepage,superscriptaddress]{revtex4-1}
\usepackage[colorlinks=true, citecolor=magenta, linkcolor=blue, urlcolor=blue]{hyperref}
\usepackage{graphicx}
\usepackage{amsmath}
\usepackage{amssymb}
\usepackage{amsfonts}
\usepackage{hyperref}
\usepackage{mathtools}
\usepackage{xcolor}

\DeclareMathOperator{\im}{Im}

\newcommand{\br}{{\bf r}}
\newcommand{\bk}{{\bf k}}
\newcommand{\bq}{{\bf q}}
\newcommand{\bs}{{\bf s}}
\newcommand{\bS}{{\bf S}}

\begin{document}
 
\title{Magnon frequency renormalization by the electronic geometrical spin torque in itinerant magnets}
\author{Emil Vi\~nas Bostr\"om}
\affiliation{Max Planck Institute for the Structure and Dynamics of Matter, Luruper Chaussee 149, 22761 Hamburg, Germany}
\email{emil.vinas-bostrom@mpsd.mpg.de}
\author{Florian G. Eich}
\affiliation{Max Planck Institute for the Structure and Dynamics of Matter, Luruper Chaussee 149, 22761 Hamburg, Germany}
\affiliation{HQS Quantum Simulations GmbH, Haid-und-Neu-Straße 7, D-76131 Karlsruhe, Germany}
\author{Angel Rubio}
\affiliation{Max Planck Institute for the Structure and Dynamics of Matter, Luruper Chaussee 149, 22761 Hamburg, Germany}
\affiliation{Center for Computational Quantum Physics (CCQ), Flatiron Institute, 162 Fifth Avenue, New York, NY 10010, USA}
\email{angel.rubio@mpsd.mpg.de}
\date{\today}

\begin{abstract}
We investigate non-adiabatic effects on the magnon frequency in an interacting system of localized spins and itinerant electrons. Including the lowest order corrections to the adiabatic dynamics in an analytically solvable model, applicable to simple ferromagnets like Fe, Co and Ni, we find that the magnon frequency is renormalized by a geometrical torque arising from the electronic spin Berry curvature. Comparison to exact numerical simulations reveals that our analytical solution captures essential low-energy features, and provides a mechanism for the magnon frequency hardening observed in recent first principles calculations for Fe, provided the geometrical torque is taken into account.
\end{abstract}

\maketitle


Understanding the dynamics of low-energy magnetic excitations is a promising route towards realizing compact and energy efficient spintronics and magnonics devices~\cite{barman_2021_2021,RMP_Tserkovnyak_2018,Jungwirth2018}. In magnetic insulators, effective spin models give a good description of the magnetic properties~\cite{Castelnovo2008,Chen2021,Suzuki2021} due to the large energy gap between charge and magnetic excitations. However, in itinerant magnets, the spin-electron coupling needs to be accounted for. For systems with large-spin ions such as Fe, Co and Ni, a semi-classical approximation of the coupled spin-electron dynamics is typically employed~\cite{Ciornei2011,Bhattacharjee2012,Cheng2016,Wadley2016,Neeraj2020}. This is seemingly justified since the large difference in time scales of electronic and magnetic excitations permits a separation into fast and slow degrees of freedom: If the electrons are assumed to follow the spins adiabatically, they can be integrated out and incorporated via additional terms in the effective spin equation of motion~\cite{Niu1998,Niu1999,Zhang06}. The effective terms are proportional to a spin Berry curvature that arises due to the adiabatic transport of the electronic states through the phase space defined by the spins. Since the momentum space Berry curvature underlies a variety of topological phenomena~\cite{Kane05,Bernevig06,Chisnell15,Chen18}, also the spin Berry curvature can be expected to have a significant influence on the system.

To calculate the magnon excitations of realistic materials, usually one out of the following two strategies is adopted: A mapping of the system onto an effective spin model, whose parameters are either computed from first principles or fitted to experimental data~\cite{Niu98,Bergqvist13}, or the use of {\it ab initio} methods such as time-dependent density functional theory (TD-DFT) to perturbatively calculate the frequency dependent transverse spin-spin response function~\cite{Savrasov98,Rousseau12,Singh19}. Although these approaches work well in many situations, the mapping to a spin model necessarily neglects dynamical effects coming from electronic excitations, and most {\it ab initio} calculations are restricted to linear response in the magnetic field strength. As recent first principles calculations have shown~\cite{TancogneDejean20}, adiabatic treatments miss qualitative effects of the magnon dynamics such as an anomalous hardening and narrowing of spectral peaks with increased perturbation strength. It is therefore of large interest to extend present treatments to address corrections to the adiabatic approximation.

Here we present a semi-classical model of magnon dynamics in itinerant magnets appropriate for simple ferromagnets like Fe, Co and Ni. The model is analytically solvable to next-to-leading order in the adiabatic parameter $\omega/\epsilon$, where $\omega$ and $\epsilon$ are the respective characteristic energy scales of the spin and electron system. We show that a careful treatment of the adiabatic limit gives rise to a geometrical torque in the effective spin equation of motion, which is due to the electronic spin Berry curvature. Including the geometrical torque in the spin equation of motion is found to have a drastic effect on the magnon frequency over a range of spin-electron couplings, which corresponds to the regime where low-energy spin-flip excitation are resonantly created in the electronic system. The effects on the magnon frequency are largest in the weak perturbation regime, and our results are thus of relevance to interpret linear response calculations. The dependence of the magnon frequency on the strength of an external perturbation is in good agreement with first principles results for Fe in the non-linear response regime~\cite{TancogneDejean20}.


We consider a coupled system of spins and itinerant electrons, appropriate for itinerant magnets such as Fe, Co and Ni, described by a Hamiltonian of the form
\begin{align}\label{eq:hamiltonian}
 H = &-t\sum_{\langle ij\rangle\sigma} \hat{c}_{i\sigma}^\dagger \hat{c}_{j\sigma} -J\sum_{\langle ij\rangle} \hat{\bS}_i\cdot\hat{\bS}_j - \sum_i (g \hat{\bs}_i + {\bf B}_i) \cdot \hat{\bS}_i.
\end{align}
Here the operator $\hat{c}_{i\sigma}$ destroys an electron at site $i$ of spin projection $\sigma$, and $\hat{\bS}_i$ is the operator at site $i$ for a localized spin of magnitude $S$. The parameters $t$ and $J$ respectively determine the nearest neighbor electron hopping amplitude and strength of the exchange interaction, $g$ gives the spin-electron coupling, and ${\bf B}_i$ is an external magnetic field. The electronic spin operator is defined by $\hat{\bs}_i = \sum_{\sigma\sigma'} \hat{c}_{i\sigma}^\dagger \boldsymbol\tau_{\sigma\sigma'} \hat{c}_{i\sigma'}$, where $\boldsymbol\tau$ is the vector of Pauli matrices. In the following we assume that the localized spin operators can be replaced by their quantum averages, $\hat{\bS}_i = \langle \hat{\bS}_i \rangle \equiv \bS_i$, which becomes exact for $S \to \infty$~\cite{Lieb73,Fradkin13}.


To derive an effective equation of motion for the spins, with the electrons in their instantaneous ground state, we consider the problem in the Lagrangian formalism. This approach was recently employed to study the dynamics of a single spin interacting with a bath~\cite{Stahl17}, and two coupled spin systems with different time scales~\cite{Elbracht2020}. Here, it is generalized to extended spin-electron systems. The Lagrangian of the coupled system is~\cite{Altland10}
\begin{align}
 \mathcal{L} &= \sum_i{\bf A}_i(\bS_i) \cdot\dot\bS_i + i\langle\Psi|\frac{\partial}{\partial t}|\Psi\rangle - \langle\Psi|H|\Psi\rangle,
\end{align}
where the vector ${\bf A}_i$ satisfies the condition $\nabla_{\bS_i} \times {\bf A}_i = -\bS_i$. In the adiabatic limit the electronic wave function depends parametrically on $\bS = \{\bS_1, \bS_2, \ldots, \bS_n\}$, and so the second term can be written $i\langle\Psi[\bS]|\partial_t|\Psi[\bS]\rangle = i\sum_i \langle\Psi[\bS]|\partial_{\bS_i} |\Psi[\bS]\rangle \cdot \dot \bS_i$. Taking the variation of the Lagrangian with respect to $\bS_i$ the spin equation of motion is
\begin{align}\label{eq:eom_omega}
 \dot{\bS}_i = \bS_i\times \bigg(J\sum_{\langle j\rangle} \bS_j + g\langle \bs_i \rangle + {\bf B}_i + {\bf T}_i[\bS]\bigg),
\end{align}
where $\bs_i = \langle \hat{\bs}_i \rangle = \langle\Psi| \hat{\bs}_i |\Psi\rangle$.
The last term on the right hand side is a geometrical torque arising from the adiabatic evolution of the electrons, and is given by
\begin{align}
 T_{i,\alpha} = -2\sum_{j\beta} \im \left(\frac{\partial\langle\Psi|}{\partial S_{i,\alpha}} \frac{\partial|\Psi\rangle}{\partial S_{j,\beta}} \right) \dot{S}_{j,\beta} = \sum_{j\beta} \Omega^{\alpha\beta}_{ij} \dot{S}_{j,\beta}.
\end{align}
Here Greek indices denote the components of a spin vector, and $\Omega_{ij}^{\alpha\beta}$ is the spin Berry curvature of the electrons. A similar equation of motion for the magnetization has previously been derived within TD-DFT~\cite{Niu98,Qian02,Niu99}.


To obtain the magnon frequency, we first calculate the electronic magnetization $\bs_i$ and the geometrical torque ${\bf T}_i[\bS]$ in the instantaneous electronic ground state. For this purpose we employ the spin spiral ansatz, i.e. assume that the localized spins are in the configuration $\bS_i = [S_\rho\cos S_{\phi,i}(t),S_\rho\sin S_{\phi,i}(t),S_z]^T$ where the cylindrical spin components are $S_\rho = S\sin\theta$, $S_z = S\cos\theta$ and $S_{\phi,i}(t) = \bq \cdot \br_i-\omega t$. The opening angle $\theta$ can be thought of as a measure of the strength of a magnetic field perturbation exciting the spin wave~\cite{TancogneDejean20}, where $\theta \to 0$ corresponds to the linear response regime.


In the following we assume for simplicity a cubic lattice of dimension $d$. For a spin spiral state the electronic Hamiltonian can be written as~\cite{Overhauser62}
\begin{align}
 H_e = \sum_\bk \Phi_\bk^\dagger
 \begin{pmatrix} \epsilon_{\bk-\bq/2} - gS_z & -gS_\rho \\ -gS_\rho & \epsilon_{\bk+\bq/2} + gS_z \end{pmatrix}
 \Phi_\bk,
\end{align}
where $\epsilon_\bk = -2t \sum_{i=1}^d \cos k_i$ is the dispersion of the bare electronic system and $\Phi_\bk = [c_{\bk-\bq/2,\uparrow}, c_{\bk+\bq/2,\downarrow}]^T$. Diagonalizing the Hamiltonian gives $H_e = \sum_{\bk s} \epsilon_{\bk s} d_{\bk s}^\dagger d_{\bk s}$ where the energies are $\epsilon_{\bk s} = (\epsilon_{\bk+\bq/2}+\epsilon_{\bk-\bq/2} + s\delta)/2$. Here $s = \pm$ and to simplify the notation we have defined $\delta^2 = (\epsilon_{\bk+\bq/2}-\epsilon_{\bk-\bq/2}+2gS_z)^2 + 4g^2S_\rho^2$. The new operators are given by a rotation of $\Phi_\bk$ with an angle $\theta_\bk = \arcsin (2gS_\rho/\delta)$. The electronic magnetization is found from the electronic density matrix and gives $\bs_i = [s_\rho \cos(\bq\cdot\br_i), s_\rho \sin(\bq\cdot\br_i), s_z]^T$, where the cylindrical components $s_\rho = (2N^d)^{-1} \sum_\bk \sin\theta_\bk (n_{\bk-} - n_{\bk+})$ and $s_z = (2N^d)^{-1} \sum_\bk \cos\theta_\bk (n_{\bk-} - n_{\bk+})$. Here $n_{\bk s}$ is the Fermi-Dirac distribution for band $s$, and we note that for general $g$ the canting angles of $\bs_i$ and $\bS_i$ are different.


\begin{figure}
 \includegraphics[width=\columnwidth]{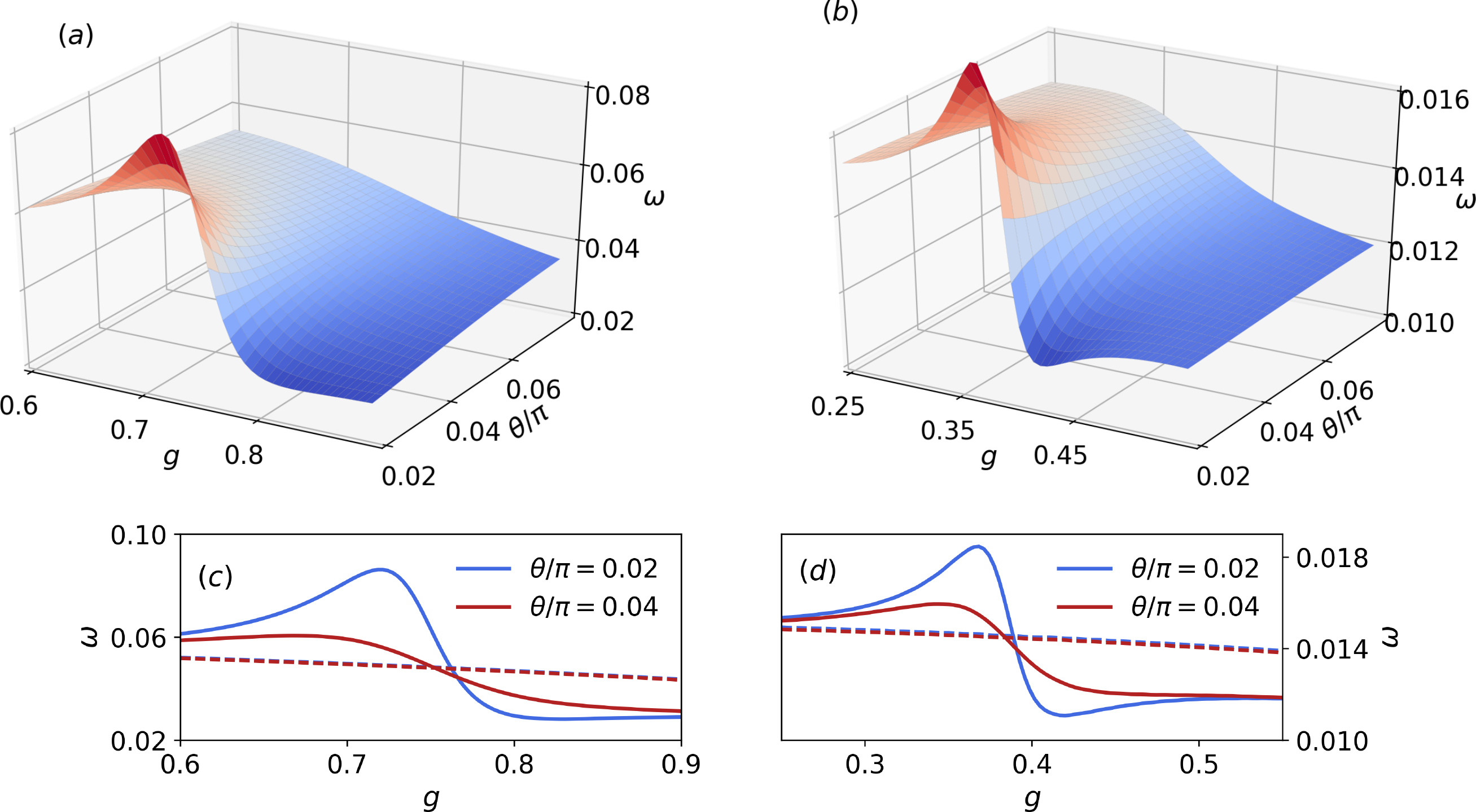}
 \caption{{\bf Magnon frequency with a geometrical torque:} $(a,b)$ Magnon frequency $\omega$ as a function of spin-electron coupling $g$ and opening angle $\theta$, for $q = \pi/4$ and $q = \pi/8$ respectively. $(c,d)$ Magnon frequency $\omega$ as a function of $g$ for $q = \pi/4$ and $q = \pi/8$ respectively, with (solid lines) and without (dashed lines) the geometrical spin torque. The results are obtained for a chain with $N = 1024$ sites, hopping amplitude $t = 1$, exchange coupling $J = 0.1$, magnetic field $B = 0$ and a spin length $S = 1$.}
 \label{fig:analytic}
\end{figure}


The Berry curvature $\Omega_{ij}$ is evaluated by writing the electronic ground state as the Slater determinant $|\Psi\rangle = \prod_{\bk s} n_{\bk s} d_{\bk s}^\dagger |0\rangle$, and calculating the derivatives of $|\Psi\rangle$ with respect to the cylindrical coordinates $\bS_i = (S_\rho,S_{\phi,i},S_z)$. This results in the tensor~\cite{SM}
\begin{align}
 \Omega_{ij}^{\alpha\beta} = \Omega^{\alpha\beta} = \begin{pmatrix} 0 & -v_z & 0 \\
                               v_z & 0 & -v_\rho \\
                               0 & v_\rho & 0
               \end{pmatrix},
\end{align}
which is independent of the site indexes $i$ and $j$ since the spin spiral is extended over the full system. The Berry curvature is written in terms of the components of the Berry potential $v_\alpha = -(1/2) \sum_{\beta\gamma}\epsilon_{\alpha\beta\gamma}\Omega^{\beta\gamma}$, given by ${\bf v} = (v_\rho,v_\phi,v_z)$. For the spiral state $v_\phi = 0$, and the remaining components have the form $v_\alpha = N^{-d} \sum_\bk v_{\alpha \bk} (n_{\bk+} - n_{\bk-})$ with $v_{\rho \bk} = g^2\sin\theta_\bk/\delta^2$ and $v_{z \bk} = g^2\cos\theta_\bk/\delta^2$. To obtain an expression for the geometrical torque ${\bf T}_i$, the Berry curvature is rotated to Cartesian coordinates where ${\bf T}_i = S_\rho\omega [v_z\cos S_{\phi,i}, v_z\sin S_{\phi,i}, -v_\rho]^T$. For later convenience we also define the spherical components $s_\theta = \cos\theta s_\rho - \sin\theta s_z$ and $v_r = \sin\theta v_\rho + \cos\theta v_z$.


Eq.~\ref{eq:eom_omega} can now be solved using the spin spiral ansatz. Due to the structure of the geometrical torque, the term proportional to ${\bf T}$ in the spin equation of motion can be equivalently represented by a frequency dependent effective magnetic field pointing along the $z$-direction. The equation of motion is therefore $\dot{\bS}_i = \bS_i\times (J\sum_{\langle j\rangle} \bS_j + B_{\text{eff}} \hat{\bf z})$, where $B_{\text{eff}} = B - g s_\theta/\sin\theta - \omega S v_r$. Solving for $\omega_\bq$ we find
\begin{align}\label{eq:frequency}
 \omega_\bq = \frac{\omega_{0\bq} - gs_{\theta,\bq}/\sin\theta}{1 + S v_{r,\bq}},
\end{align}
where $\omega_{0\bq} = 4JS\cos\theta \sum_{i=1}^d \sin^2(q_i/2) + B$ is the magnon frequency of the isolated spin system. This equation is the main result of this paper, and gives the magnon frequency in the semi-adiabatic limit in presence of the geometrical torque.

\begin{figure}
 \includegraphics[width=\columnwidth]{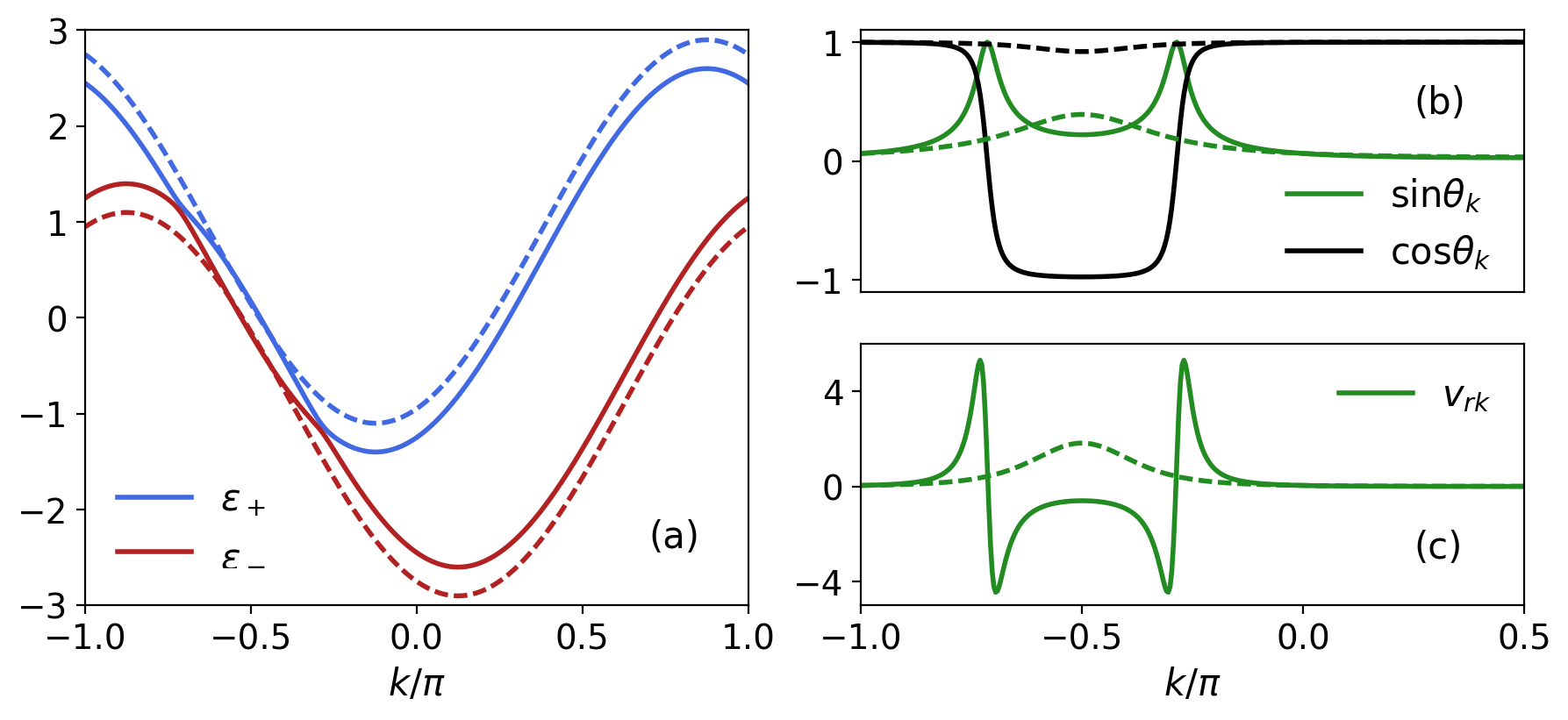}
 \caption{{\bf Electronic ground state properties:} $(a)$ Electronic dispersion $\epsilon_{ks}$, $(b)$ mixing angles $\sin\theta_k$ and $\cos\theta_k$,  and $(c)$ radial Berry potential $v_{rk}$ as a function of wave vector $k$. Solid (dashed) lines correspond to an electron-spin coupling $g = 0.6$ ($g = 0.9$). The parameters are as in Fig.~\ref{fig:analytic} with $q = \pi/4$.}
 \label{fig:parameters}
\end{figure}

Although Eq.~\ref{eq:frequency} is valid in arbitrary dimensions, we restrict in the following to $d = 1$ where exact numerical comparisons are easily obtained. Fig.~\ref{fig:analytic} shows the magnon frequency as a function of the spin-electron coupling $g$ and the spiral opening angle $\theta$, for wave vectors $q = \pi/4$ and $q = \pi/8$. Both cases display a regime of coupling strengths where $\omega$ is an increasing function of $\theta$, starting from a critical value $g = g_c$. This anomalous hardening is restricted to a finite range, and for $g$ much smaller or larger than $g_c$ the frequency softens with increasing $\theta$. The value of $g_c$ is empirically identified as the coupling for which $\Delta\epsilon = \epsilon_{k+q/2}-\epsilon_{k-q/2} - 2gS_z \approx 0$, which is the value of $g$ where the electronic bands become disentangled (see Fig.~\ref{fig:parameters}). For $g < g_c$ the difference $\delta = \epsilon_{k+} - \epsilon_{k-}$ has two minima symmetrically located around $k = -\pi/2$, that with increasing $g$ come together until they merge at $k = -\pi/2$ for $g \approx g_c$. The value of $g_c$ can thus be estimated by requiring that $\Delta\epsilon = 0$, which gives $g_c \approx 2\sin(q/2)$. The resulting values $g_c \approx 0.40$ for $q = \pi/8$ and $g_c \approx 0.76$ for $q = \pi/4$ are in good agreement with the numerical results of Fig.~\ref{fig:analytic}, and indicate that the anomalous behavior of $\omega$ appears for larger $g$ with increasing $q$.

To understand the origin of the anomalous hardening, we compare the frequencies with and without the geometrical torque (Figs.~\ref{fig:analytic}c and \ref{fig:analytic}d). We note that the hardening vanishes for ${\bf T} = 0$, and that the main effect of the geometrical torque is seen at small $\theta$. A likely explanation of the observed trends is that the electrons provide a dynamical torque through the creation of low-energy excitations. This is corroborated by noting that the expression $\Delta\epsilon = \epsilon_{k+q/2}-\epsilon_{k-q/2} + 2gS_z$ also appears as the denominator of the transverse electronic spin-spin response function~\cite{Eich13}, so that the condition $\Delta\epsilon \approx 0$ is equivalent to a resonant response of the electronic system. In analogy with the non-adiabatic couplings used to induce transitions at conical intersections of Born-Oppenheimer surfaces~\cite{Zhang06,Baer06}, the action of the geometrical torque can be interpreted as inducing transitions between the energy surfaces $\epsilon_{ks}$. We note that for fixed $q$ and $k \approx -\pi/2$, the energy of spin-flip excitations $\delta$ and the non-adiabatic coupling $v_{rk}$ are increasing and decreasing functions of $\theta$, respectively. Since non-adiabatic transitions should be favored by a small spin-flip gap and large non-adiabatic coupling, the electronic response is expected to be strongest for small $\theta$ in line with the observed trends.


In the non-interacting limit, i.e. for $g = 0$, Eq.~\ref{eq:frequency} reduces to the frequency $\omega_{0\bq}$ of an isolated Heisenberg system and is a monotonically decreasing function of $\theta$. In the opposite limit, where $gS \gg \epsilon_k$, the expressions for $\bs$ and ${\bf v}$ can be worked out analytically by noting that $\delta \approx 2gS$, $\sin\theta_k \approx S_\rho/S$ and $\cos\theta_k \approx S_z/S$. It follows that the only non-zero components of the magnetization and Berry potential are $s_r = (k_-^F - k_+^F)/2\pi$ and $v_r = -(k_-^F - k_+^F)/4\pi S^2$. The Berry potential can thus be written in terms of the magnitude $M_s = |\bs|$ of the electronic magnetization like $v_r = -M_s/2S^2$. If we denote the magnitude of the electron spin by $s = 1/2$ the expression for $v_r$ becomes identical to that of a single localized spin, $v_r = -M_s s/S^2$~\cite{Stahl17}. In the strong coupling limit the frequency is thus given by $\omega = \omega_0/(1-M_s s/S)$.

\begin{figure}
 \includegraphics[width=\columnwidth]{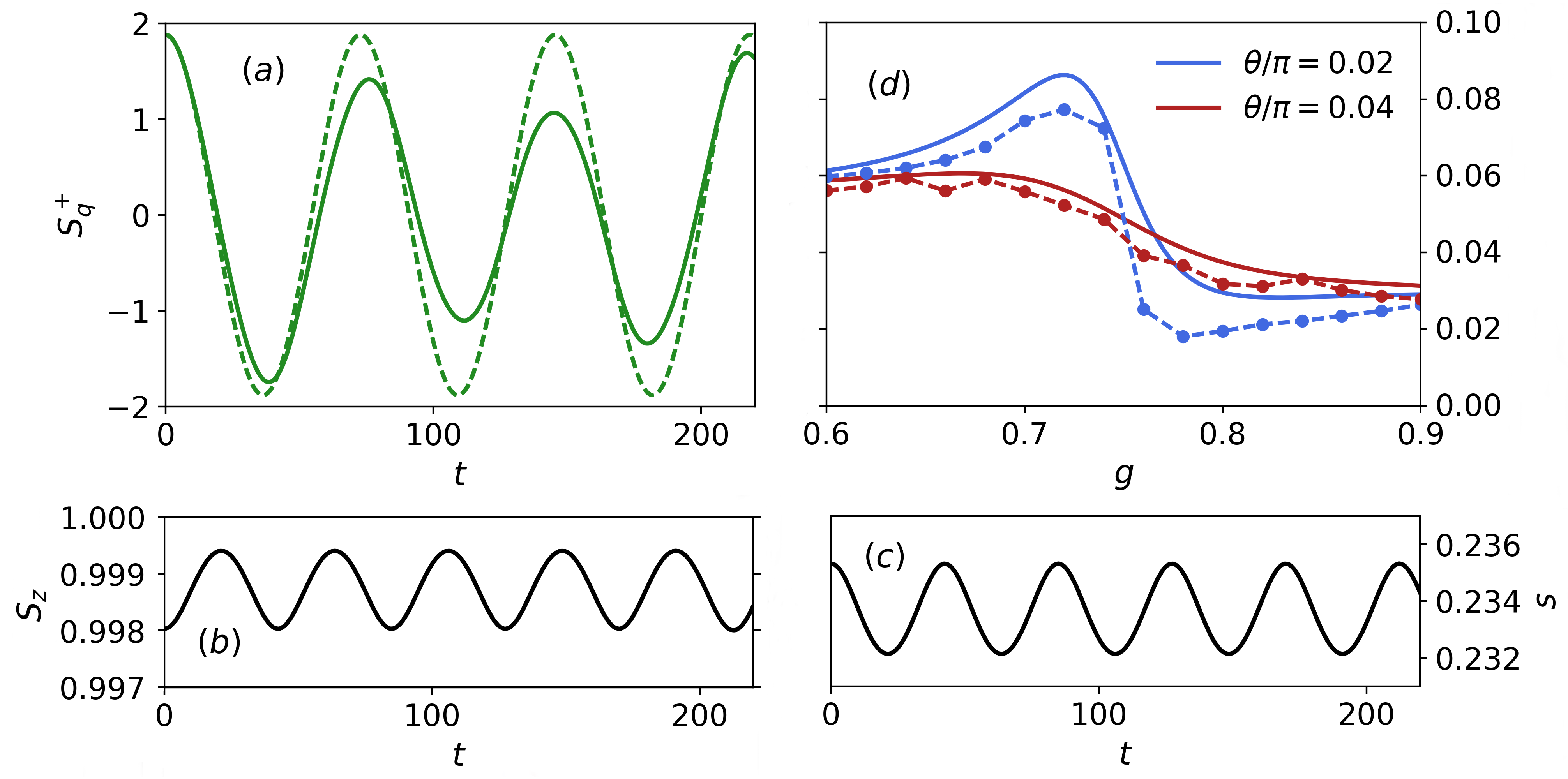}
 \caption{{\bf Comparison of numerical and analytical results:} $(a)$ Time signal $S^+_q(t)$ for semi-adiabatic (dashed) and non-adiabatic evolution (solid). $(b,c)$ The $z$-component of the localized spins $S_z$ and the magnitude $s = |\bs|$ of the electronic magnetization. $(d)$ Magnon frequency $\omega$ as a function of spin-electron coupling $g$. Solid (dashed) lines show the non-adiabatic (semi-adiabatic) results. The parameters are as in Fig.~\ref{fig:analytic} with $q = \pi/4$.}
 \label{fig:numeric}
\end{figure}

For large $S$ the denominator tends to unity, and the frequency reduces to $\omega_0$. Instead, in the strong coupling limit $g \to \infty$ with $S$ finite, the frequency is uniformly renormalized by the Berry potential of the electronic system: since $M_s$ is independent of $\theta$ in the large coupling limit, the renormalization does not affect the $q$- or $\theta$-dependence of the frequency. This indicates that the relative canting of the itinerant and isolated spins, present for intermediate coupling strengths, is necessary to alter the dependence of $\omega$ on $\theta$. We note that for $g \to \infty$ the magnetization $M_s = 1$, and the spin Chern number $\mathcal{C} = (2\pi)^{-1} \int d\bS \cdot {\bf v} = 1$. In contrast, the limit $g = 0$ corresponds to a spin Chern number $\mathcal{C} = 0$. Thus, the electronic system undergoes a topological transition as a function of $g$. Since the Chern number is only quantized in gapped systems, we identify the transition point as the value $g = 2$ where the system becomes insulating.


To assess the validity of the semi-adiabatic approximation, the magnon frequency of Eq.~\ref{eq:frequency} was compared to a numerical solution of Eq.~\ref{eq:hamiltonian} assuming classical spins. Starting from the electronic ground state in presence of a spin spiral, the electronic system was propagated using the short iterated Lanczos algorithm~\cite{Park86} and the spin system using the Depondt-Mertens algorithm~\cite{Depondt09}, coupled using Heun's predictor-corrector method. The magnon frequency was extracted from the time signal $S_k^+(t) = \sum_i e^{-ikr_i} (S_i^x(t) + iS_i^y(t))$, which was found to vanish for all $k \neq q$. The signal is well described by a function of the form $S_q^+(t) = A e^{i(qr_i - \omega_1 t)} + B e^{i(qr_i - \omega_2 t)}$, where $B \approx 0$ except for $g \approx g_c$. By fitting this function to the numerical time signal we obtained the fundamental frequency $\omega$ for given $g$ and $\theta$ without the extensive time propagation needed for a Fourier transform.

Away from the region $g \approx g_c$ the the semi-adiabatic result is in good quantitative agreement with numerical frequency, as seen in Fig.~\ref{fig:numeric}d. The additional oscillations in the numerical results are a finite size effect arising from the $k$-point sampling in the region around $k = -\pi/2$, where both the mixing angle $\theta_k$ and the Berry curvature $\Omega$ changes rapidly (see Fig.~\ref{fig:parameters}). We have checked that the oscillations diminish in strength with increasing system size. In the region $g \approx g_c$ the semi-adiabatic theory still gives a good qualitative description of the frequency, with the main difference being that the numerical results show a larger and more rapid drop in $\omega$ when $g$ crosses $g_c$.

When two modes are present in $S_q^+(t)$ there is an associated nutational motion of the localized spins and a modulation of the magnitude of the electronic magnetization, both with a frequency $\omega = \omega_1 - \omega_2$ (see Fig.~\ref{fig:numeric}b and \ref{fig:numeric}c). For $g < g_c$ the $\omega_2$ mode propagates in opposite direction to the $\omega_1$ mode, while for $g > g_c$ they propagate in the same direction. This is understood by noting that for $g < g_c$ the operator $d_+^\dagger d_- \approx c_{k-q/2,\uparrow}^\dagger c_{k+q/2,\downarrow}$ responsible electronic spin-flip excitations corresponds to an excitation of energy $\delta$ and momentum $q$, while for $g > g_c$ the operator $d_+^\dagger d_- \approx c_{k+q/2,\downarrow}^\dagger c_{k-q/2,\uparrow}$ creates an excitation of energy $\delta$ and momentum $-q$. Thus the mode $\omega_2$ can be associated with spin-flip excitations in the electronic system, and indeed numerically we find $\omega_2 \approx \delta$.


\begin{figure}
 \includegraphics[width=\columnwidth]{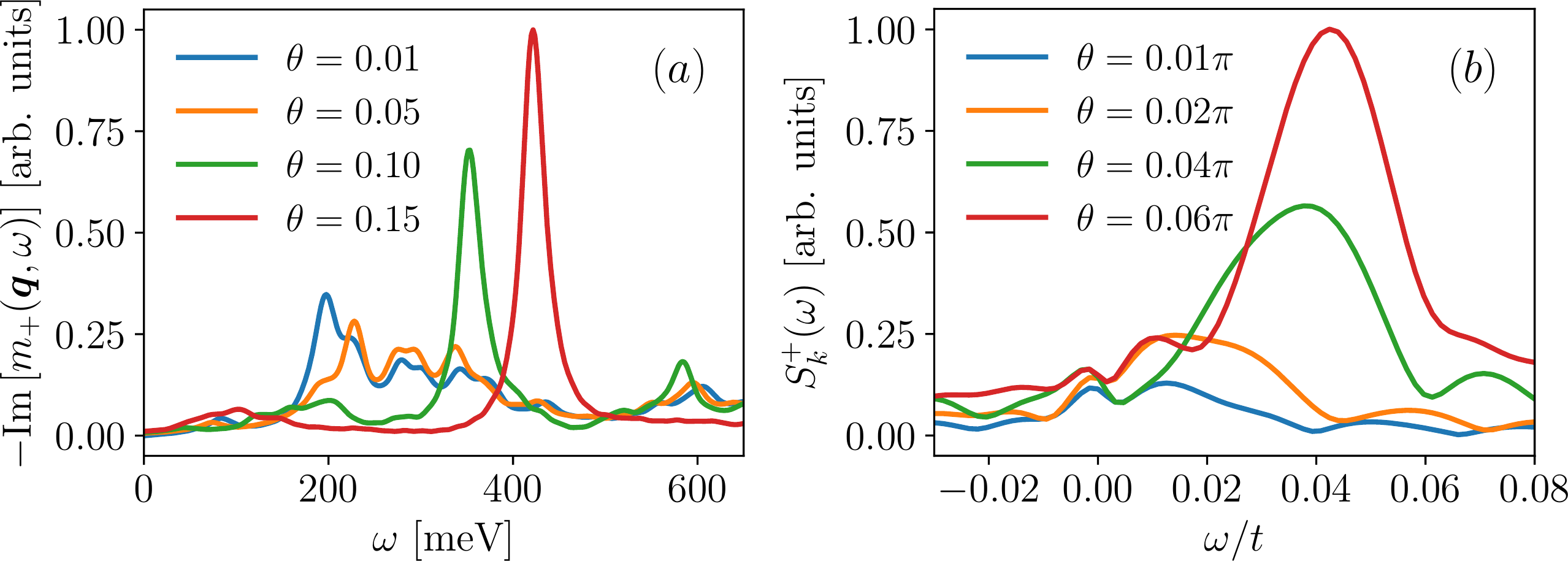}
 \caption{{\bf Comparison of model and TD-DFT results:} $(a)$ Magnon density of states of BCC Fe for $\bq = \tfrac{1}{4} (-1,1,1)$, obtained from a real-time calculation of the magnetic susceptibility within TD-DFT~\cite{TancogneDejean20}. $(b)$ Fourier transform of the time signal $S^+_q(t)$ from time-evolving the model of Eq.~\ref{eq:hamiltonian} with classical spins. The parameters are as in Fig.~\ref{fig:analytic}, with $g = 0.78$ and $q = \pi/4$.}
 \label{fig:tddft}
\end{figure}


Although the semi-adiabatic approach fails to capture the presence of a second mode for $g \approx g_c$, it gives a good description of the fundamental frequency over the full range of $g$. In particular, the semi-adiabatic result correctly captures the anomalous hardening of $\omega$ when $g > g_c$. This effect is due solely to the presence of the geometrical torque in the spin equation of motion, which introduces leading order non-adiabatic effects into the otherwise adiabatic description. We note that since the rate of spin-flip excitations is expected to diminish with $\theta$ (because of the decrease of $v_r$ and increase of $\delta$), the magnon lifetime should be a increasing function of $\theta$ in the region $g \approx g_c$. This leads to sharper spectral peaks with increasing $\theta$, and thus both the anomalous hardening and narrowing of the spectral peaks are in agreement with recent first principles calculations of Fe~\cite{TancogneDejean20}.

In Fig.~\ref{fig:tddft} we compare the behaviour of the magnon frequency of bulk Fe found via real-time calculations of the magnetic susceptibility with TD-DFT~\cite{TancogneDejean20}, to the results obtain by a numerical solution of Eq.~\ref{eq:hamiltonian}. We see that the qualitative features, namely the shift and narrowing of the spectral peaks with $\theta$, are in agreement. In contrast, TD-DFT calculations for the antiferromagnetic insulator NiO shows a softening of the magnon frequency with increasing $\theta$ as expected from a pure spin model~\cite{Nicolas}. Since the model has a very simple band structure compared to bulk Fe, this indicates the generality of the presented results. We note that the critical value of the spin-electron coupling where the anomalous behavior is strongest depends on $q$ approximately as $g_c \approx 2\sin(q/2)$, and therefore it is likely that some magnons are affected by the geometrical torque in a large class of magnetic materials.

In conclusion, the semi-adiabatic spin dynamics presented provides an intuitive understanding of the dynamics of coupled spin-electron systems, where non-adiabatic effects due to the geometrical torque arise from the spin Berry curvature of the electrons. Via the Berry curvature the electrons exert a dynamical torque through the creation of low-energy spin-flip excitations. Since our results only depend on the presence of a non-zero Berry curvature in the spin equation of motion, we expect similar effects to appear for other coupled systems of fast and slow degrees of freedom, and to be a generic feature of the spin dynamics of itinerant magnets.


\begin{acknowledgments} 
{\it Acknowledgments.-} We acknowledge inspiring discussions with Nicolas Tancogne-Dejean. We acknowledge support by the Max Planck Institute - New York City Center for Non-Equilibrium Quantum Phenomena. This work was supported by the European Research Council (ERC-2015-AdG694097), the Cluster of Excellence “Advanced Imaging of Matter” (AIM), and Grupos Consolidados (IT1249-19). The Flatiron Institute is a Division of the Simons Foundation.
\end{acknowledgments}


\bibliography{references}

\begin{thebibliography}{38}%
\makeatletter
\providecommand \@ifxundefined [1]{%
 \@ifx{#1\undefined}
}%
\providecommand \@ifnum [1]{%
 \ifnum #1\expandafter \@firstoftwo
 \else \expandafter \@secondoftwo
 \fi
}%
\providecommand \@ifx [1]{%
 \ifx #1\expandafter \@firstoftwo
 \else \expandafter \@secondoftwo
 \fi
}%
\providecommand \natexlab [1]{#1}%
\providecommand \enquote  [1]{``#1''}%
\providecommand \bibnamefont  [1]{#1}%
\providecommand \bibfnamefont [1]{#1}%
\providecommand \citenamefont [1]{#1}%
\providecommand \href@noop [0]{\@secondoftwo}%
\providecommand \href [0]{\begingroup \@sanitize@url \@href}%
\providecommand \@href[1]{\@@startlink{#1}\@@href}%
\providecommand \@@href[1]{\endgroup#1\@@endlink}%
\providecommand \@sanitize@url [0]{\catcode `\\12\catcode `\$12\catcode
  `\&12\catcode `\#12\catcode `\^12\catcode `\_12\catcode `\%12\relax}%
\providecommand \@@startlink[1]{}%
\providecommand \@@endlink[0]{}%
\providecommand \url  [0]{\begingroup\@sanitize@url \@url }%
\providecommand \@url [1]{\endgroup\@href {#1}{\urlprefix }}%
\providecommand \urlprefix  [0]{URL }%
\providecommand \Eprint [0]{\href }%
\providecommand \doibase [0]{http://dx.doi.org/}%
\providecommand \selectlanguage [0]{\@gobble}%
\providecommand \bibinfo  [0]{\@secondoftwo}%
\providecommand \bibfield  [0]{\@secondoftwo}%
\providecommand \translation [1]{[#1]}%
\providecommand \BibitemOpen [0]{}%
\providecommand \bibitemStop [0]{}%
\providecommand \bibitemNoStop [0]{.\EOS\space}%
\providecommand \EOS [0]{\spacefactor3000\relax}%
\providecommand \BibitemShut  [1]{\csname bibitem#1\endcsname}%
\let\auto@bib@innerbib\@empty
\bibitem [{\citenamefont {Barman}\ \emph {et~al.}(2021)\citenamefont {Barman},
  \citenamefont {Gubbiotti}, \citenamefont {Ladak}, \citenamefont {Adeyeye},
  \citenamefont {Krawczyk}, \citenamefont {Gr{\"a}fe}, \citenamefont
  {Adelmann}, \citenamefont {Cotofana}, \citenamefont {Naeemi}, \citenamefont
  {Vasyuchka}, \citenamefont {Hillebrands}, \citenamefont {Nikitov},
  \citenamefont {Yu}, \citenamefont {Grundler}, \citenamefont {Sadovnikov},
  \citenamefont {Grachev}, \citenamefont {Sheshukova}, \citenamefont
  {Duquesne}, \citenamefont {Marangolo}, \citenamefont {Gyorgy}, \citenamefont
  {Porod}, \citenamefont {Demidov}, \citenamefont {Urazhdin}, \citenamefont
  {Demokritov}, \citenamefont {Albisetti}, \citenamefont {Petti}, \citenamefont
  {Bertacco}, \citenamefont {Schulteiss}, \citenamefont {Kruglyak},
  \citenamefont {Poimanov}, \citenamefont {Sahoo}, \citenamefont {Sinha},
  \citenamefont {Yang}, \citenamefont {Muenzenberg}, \citenamefont {Moriyama},
  \citenamefont {Mizukami}, \citenamefont {Landeros}, \citenamefont {Gallardo},
  \citenamefont {Carlotti}, \citenamefont {Kim}, \citenamefont {Stamps},
  \citenamefont {Camley}, \citenamefont {Rana}, \citenamefont {Otani},
  \citenamefont {Yu}, \citenamefont {Yu}, \citenamefont {Bauer}, \citenamefont
  {Back}, \citenamefont {Uhrig}, \citenamefont {Dobrovolskiy}, \citenamefont
  {Dijken}, \citenamefont {Budinska}, \citenamefont {Qin}, \citenamefont
  {Chumak}, \citenamefont {Khitun}, \citenamefont {Nikonov}, \citenamefont
  {Young}, \citenamefont {Zingsem},\ and\ \citenamefont
  {Winklhofer}}]{barman_2021_2021}%
  \BibitemOpen
  \bibfield  {author} {\bibinfo {author} {\bibfnamefont {A.}~\bibnamefont
  {Barman}}, \bibinfo {author} {\bibfnamefont {G.}~\bibnamefont {Gubbiotti}},
  \bibinfo {author} {\bibfnamefont {S.}~\bibnamefont {Ladak}}, \bibinfo
  {author} {\bibfnamefont {A.~O.}\ \bibnamefont {Adeyeye}}, \bibinfo {author}
  {\bibfnamefont {M.}~\bibnamefont {Krawczyk}}, \bibinfo {author}
  {\bibfnamefont {J.}~\bibnamefont {Gr{\"a}fe}}, \bibinfo {author}
  {\bibfnamefont {C.}~\bibnamefont {Adelmann}}, \bibinfo {author}
  {\bibfnamefont {S.}~\bibnamefont {Cotofana}}, \bibinfo {author}
  {\bibfnamefont {A.}~\bibnamefont {Naeemi}}, \bibinfo {author} {\bibfnamefont
  {V.~I.}\ \bibnamefont {Vasyuchka}}, \bibinfo {author} {\bibfnamefont
  {B.}~\bibnamefont {Hillebrands}}, \bibinfo {author} {\bibfnamefont {S.~A.}\
  \bibnamefont {Nikitov}}, \bibinfo {author} {\bibfnamefont {H.}~\bibnamefont
  {Yu}}, \bibinfo {author} {\bibfnamefont {D.}~\bibnamefont {Grundler}},
  \bibinfo {author} {\bibfnamefont {A.}~\bibnamefont {Sadovnikov}}, \bibinfo
  {author} {\bibfnamefont {A.~A.}\ \bibnamefont {Grachev}}, \bibinfo {author}
  {\bibfnamefont {S.~E.}\ \bibnamefont {Sheshukova}}, \bibinfo {author}
  {\bibfnamefont {J.-Y.}\ \bibnamefont {Duquesne}}, \bibinfo {author}
  {\bibfnamefont {M.}~\bibnamefont {Marangolo}}, \bibinfo {author}
  {\bibfnamefont {C.}~\bibnamefont {Gyorgy}}, \bibinfo {author} {\bibfnamefont
  {W.}~\bibnamefont {Porod}}, \bibinfo {author} {\bibfnamefont {V.~E.}\
  \bibnamefont {Demidov}}, \bibinfo {author} {\bibfnamefont {S.}~\bibnamefont
  {Urazhdin}}, \bibinfo {author} {\bibfnamefont {S.}~\bibnamefont
  {Demokritov}}, \bibinfo {author} {\bibfnamefont {E.}~\bibnamefont
  {Albisetti}}, \bibinfo {author} {\bibfnamefont {D.}~\bibnamefont {Petti}},
  \bibinfo {author} {\bibfnamefont {R.}~\bibnamefont {Bertacco}}, \bibinfo
  {author} {\bibfnamefont {H.}~\bibnamefont {Schulteiss}}, \bibinfo {author}
  {\bibfnamefont {V.~V.}\ \bibnamefont {Kruglyak}}, \bibinfo {author}
  {\bibfnamefont {V.~D.}\ \bibnamefont {Poimanov}}, \bibinfo {author}
  {\bibfnamefont {A.~K.}\ \bibnamefont {Sahoo}}, \bibinfo {author}
  {\bibfnamefont {J.}~\bibnamefont {Sinha}}, \bibinfo {author} {\bibfnamefont
  {H.}~\bibnamefont {Yang}}, \bibinfo {author} {\bibfnamefont {M.}~\bibnamefont
  {Muenzenberg}}, \bibinfo {author} {\bibfnamefont {T.}~\bibnamefont
  {Moriyama}}, \bibinfo {author} {\bibfnamefont {S.}~\bibnamefont {Mizukami}},
  \bibinfo {author} {\bibfnamefont {P.}~\bibnamefont {Landeros}}, \bibinfo
  {author} {\bibfnamefont {R.~A.}\ \bibnamefont {Gallardo}}, \bibinfo {author}
  {\bibfnamefont {G.}~\bibnamefont {Carlotti}}, \bibinfo {author}
  {\bibfnamefont {J.-V.}\ \bibnamefont {Kim}}, \bibinfo {author} {\bibfnamefont
  {R.~L.}\ \bibnamefont {Stamps}}, \bibinfo {author} {\bibfnamefont {R.~E.}\
  \bibnamefont {Camley}}, \bibinfo {author} {\bibfnamefont {B.}~\bibnamefont
  {Rana}}, \bibinfo {author} {\bibfnamefont {Y.}~\bibnamefont {Otani}},
  \bibinfo {author} {\bibfnamefont {W.}~\bibnamefont {Yu}}, \bibinfo {author}
  {\bibfnamefont {T.}~\bibnamefont {Yu}}, \bibinfo {author} {\bibfnamefont
  {G.~E.~W.}\ \bibnamefont {Bauer}}, \bibinfo {author} {\bibfnamefont {C.~H.}\
  \bibnamefont {Back}}, \bibinfo {author} {\bibfnamefont {G.~S.}\ \bibnamefont
  {Uhrig}}, \bibinfo {author} {\bibfnamefont {O.~V.}\ \bibnamefont
  {Dobrovolskiy}}, \bibinfo {author} {\bibfnamefont {S.~v.}\ \bibnamefont
  {Dijken}}, \bibinfo {author} {\bibfnamefont {B.}~\bibnamefont {Budinska}},
  \bibinfo {author} {\bibfnamefont {H.}~\bibnamefont {Qin}}, \bibinfo {author}
  {\bibfnamefont {A.}~\bibnamefont {Chumak}}, \bibinfo {author} {\bibfnamefont
  {A.}~\bibnamefont {Khitun}}, \bibinfo {author} {\bibfnamefont {D.~E.}\
  \bibnamefont {Nikonov}}, \bibinfo {author} {\bibfnamefont {I.~A.}\
  \bibnamefont {Young}}, \bibinfo {author} {\bibfnamefont {B.}~\bibnamefont
  {Zingsem}}, \ and\ \bibinfo {author} {\bibfnamefont {M.}~\bibnamefont
  {Winklhofer}},\ }\href {\doibase 10.1088/1361-648X/abec1a} {\bibfield
  {journal} {\bibinfo  {journal} {Journal of Physics: Condensed Matter}\ }
  (\bibinfo {year} {2021}),\ 10.1088/1361-648X/abec1a}\BibitemShut {NoStop}%
\bibitem [{\citenamefont {Baltz}\ \emph {et~al.}(2018)\citenamefont {Baltz},
  \citenamefont {Manchon}, \citenamefont {Tsoi}, \citenamefont {Moriyama},
  \citenamefont {Ono},\ and\ \citenamefont
  {Tserkovnyak}}]{RMP_Tserkovnyak_2018}%
  \BibitemOpen
  \bibfield  {author} {\bibinfo {author} {\bibfnamefont {V.}~\bibnamefont
  {Baltz}}, \bibinfo {author} {\bibfnamefont {A.}~\bibnamefont {Manchon}},
  \bibinfo {author} {\bibfnamefont {M.}~\bibnamefont {Tsoi}}, \bibinfo {author}
  {\bibfnamefont {T.}~\bibnamefont {Moriyama}}, \bibinfo {author}
  {\bibfnamefont {T.}~\bibnamefont {Ono}}, \ and\ \bibinfo {author}
  {\bibfnamefont {Y.}~\bibnamefont {Tserkovnyak}},\ }\href {\doibase
  10.1103/RevModPhys.90.015005} {\bibfield  {journal} {\bibinfo  {journal}
  {Rev. Mod. Phys.}\ }\textbf {\bibinfo {volume} {90}},\ \bibinfo {pages}
  {015005} (\bibinfo {year} {2018})}\BibitemShut {NoStop}%
\bibitem [{\citenamefont {Jungwirth}\ \emph {et~al.}(2018)\citenamefont
  {Jungwirth}, \citenamefont {Sinova}, \citenamefont {Manchon}, \citenamefont
  {Marti}, \citenamefont {Wunderlich},\ and\ \citenamefont
  {Felser}}]{Jungwirth2018}%
  \BibitemOpen
  \bibfield  {author} {\bibinfo {author} {\bibfnamefont {T.}~\bibnamefont
  {Jungwirth}}, \bibinfo {author} {\bibfnamefont {J.}~\bibnamefont {Sinova}},
  \bibinfo {author} {\bibfnamefont {A.}~\bibnamefont {Manchon}}, \bibinfo
  {author} {\bibfnamefont {X.}~\bibnamefont {Marti}}, \bibinfo {author}
  {\bibfnamefont {J.}~\bibnamefont {Wunderlich}}, \ and\ \bibinfo {author}
  {\bibfnamefont {C.}~\bibnamefont {Felser}},\ }\href {\doibase
  10.1038/s41567-018-0063-6} {\bibfield  {journal} {\bibinfo  {journal} {Nature
  Physics}\ }\textbf {\bibinfo {volume} {14}},\ \bibinfo {pages} {200}
  (\bibinfo {year} {2018})}\BibitemShut {NoStop}%
\bibitem [{\citenamefont {Castelnovo}\ \emph {et~al.}(2008)\citenamefont
  {Castelnovo}, \citenamefont {Moessner},\ and\ \citenamefont
  {Sondhi}}]{Castelnovo2008}%
  \BibitemOpen
  \bibfield  {author} {\bibinfo {author} {\bibfnamefont {C.}~\bibnamefont
  {Castelnovo}}, \bibinfo {author} {\bibfnamefont {R.}~\bibnamefont
  {Moessner}}, \ and\ \bibinfo {author} {\bibfnamefont {S.~L.}\ \bibnamefont
  {Sondhi}},\ }\href {\doibase 10.1038/nature06433} {\bibfield  {journal}
  {\bibinfo  {journal} {Nature}\ }\textbf {\bibinfo {volume} {451}},\ \bibinfo
  {pages} {42} (\bibinfo {year} {2008})}\BibitemShut {NoStop}%
\bibitem [{\citenamefont {Chen}\ \emph {et~al.}(2021)\citenamefont {Chen},
  \citenamefont {Chung}, \citenamefont {Stone}, \citenamefont {Kolesnikov},
  \citenamefont {Winn}, \citenamefont {Garlea}, \citenamefont {Abernathy},
  \citenamefont {Gao}, \citenamefont {Augustin}, \citenamefont {Santos},\ and\
  \citenamefont {Dai}}]{Chen2021}%
  \BibitemOpen
  \bibfield  {author} {\bibinfo {author} {\bibfnamefont {L.}~\bibnamefont
  {Chen}}, \bibinfo {author} {\bibfnamefont {J.-H.}\ \bibnamefont {Chung}},
  \bibinfo {author} {\bibfnamefont {M.~B.}\ \bibnamefont {Stone}}, \bibinfo
  {author} {\bibfnamefont {A.~I.}\ \bibnamefont {Kolesnikov}}, \bibinfo
  {author} {\bibfnamefont {B.}~\bibnamefont {Winn}}, \bibinfo {author}
  {\bibfnamefont {V.~O.}\ \bibnamefont {Garlea}}, \bibinfo {author}
  {\bibfnamefont {D.~L.}\ \bibnamefont {Abernathy}}, \bibinfo {author}
  {\bibfnamefont {B.}~\bibnamefont {Gao}}, \bibinfo {author} {\bibfnamefont
  {M.}~\bibnamefont {Augustin}}, \bibinfo {author} {\bibfnamefont {E.~J.~G.}\
  \bibnamefont {Santos}}, \ and\ \bibinfo {author} {\bibfnamefont
  {P.}~\bibnamefont {Dai}},\ }\href {\doibase 10.1103/PhysRevX.11.031047}
  {\bibfield  {journal} {\bibinfo  {journal} {Phys. Rev. X}\ }\textbf {\bibinfo
  {volume} {11}},\ \bibinfo {pages} {031047} (\bibinfo {year}
  {2021})}\BibitemShut {NoStop}%
\bibitem [{Suz(2021)}]{Suzuki2021}%
  \BibitemOpen
  \href {\doibase 10.1038/s41467-021-24722-4} {\bibfield  {journal} {\bibinfo
  {journal} {Nature Comm.}\ } (\bibinfo {year} {2021}),\
  10.1038/s41467-021-24722-4}\BibitemShut {NoStop}%
\bibitem [{\citenamefont {Ciornei}\ \emph {et~al.}(2011)\citenamefont
  {Ciornei}, \citenamefont {Rub\'{\i}},\ and\ \citenamefont
  {Wegrowe}}]{Ciornei2011}%
  \BibitemOpen
  \bibfield  {author} {\bibinfo {author} {\bibfnamefont {M.-C.}\ \bibnamefont
  {Ciornei}}, \bibinfo {author} {\bibfnamefont {J.~M.}\ \bibnamefont
  {Rub\'{\i}}}, \ and\ \bibinfo {author} {\bibfnamefont {J.-E.}\ \bibnamefont
  {Wegrowe}},\ }\href {\doibase 10.1103/PhysRevB.83.020410} {\bibfield
  {journal} {\bibinfo  {journal} {Phys. Rev. B}\ }\textbf {\bibinfo {volume}
  {83}},\ \bibinfo {pages} {020410(R)} (\bibinfo {year} {2011})}\BibitemShut
  {NoStop}%
\bibitem [{\citenamefont {Bhattacharjee}\ \emph {et~al.}(2012)\citenamefont
  {Bhattacharjee}, \citenamefont {Nordstr\"om},\ and\ \citenamefont
  {Fransson}}]{Bhattacharjee2012}%
  \BibitemOpen
  \bibfield  {author} {\bibinfo {author} {\bibfnamefont {S.}~\bibnamefont
  {Bhattacharjee}}, \bibinfo {author} {\bibfnamefont {L.}~\bibnamefont
  {Nordstr\"om}}, \ and\ \bibinfo {author} {\bibfnamefont {J.}~\bibnamefont
  {Fransson}},\ }\href {\doibase 10.1103/PhysRevLett.108.057204} {\bibfield
  {journal} {\bibinfo  {journal} {Phys. Rev. Lett.}\ }\textbf {\bibinfo
  {volume} {108}},\ \bibinfo {pages} {057204} (\bibinfo {year}
  {2012})}\BibitemShut {NoStop}%
\bibitem [{\citenamefont {Cheng}\ \emph {et~al.}(2016)\citenamefont {Cheng},
  \citenamefont {Xiao},\ and\ \citenamefont {Brataas}}]{Cheng2016}%
  \BibitemOpen
  \bibfield  {author} {\bibinfo {author} {\bibfnamefont {R.}~\bibnamefont
  {Cheng}}, \bibinfo {author} {\bibfnamefont {D.}~\bibnamefont {Xiao}}, \ and\
  \bibinfo {author} {\bibfnamefont {A.}~\bibnamefont {Brataas}},\ }\href
  {\doibase 10.1103/PhysRevLett.116.207603} {\bibfield  {journal} {\bibinfo
  {journal} {Phys. Rev. Lett.}\ }\textbf {\bibinfo {volume} {116}},\ \bibinfo
  {pages} {207603} (\bibinfo {year} {2016})}\BibitemShut {NoStop}%
\bibitem [{\citenamefont {Wadley}\ \emph {et~al.}(2016)\citenamefont {Wadley},
  \citenamefont {Howells}, \citenamefont {{\v{Z}}elezn{\'{y}}}, \citenamefont
  {Andrews}, \citenamefont {Hills}, \citenamefont {Campion}, \citenamefont
  {Nov{\'{a}}k}, \citenamefont {Olejn{\'{\i}}k}, \citenamefont {Maccherozzi},
  \citenamefont {Dhesi}, \citenamefont {Martin}, \citenamefont {Wagner},
  \citenamefont {Wunderlich}, \citenamefont {Freimuth}, \citenamefont
  {Mokrousov}, \citenamefont {Kune{\v{s}}}, \citenamefont {Chauhan},
  \citenamefont {Grzybowski}, \citenamefont {Rushforth}, \citenamefont
  {Edmonds}, \citenamefont {Gallagher},\ and\ \citenamefont
  {Jungwirth}}]{Wadley2016}%
  \BibitemOpen
  \bibfield  {author} {\bibinfo {author} {\bibfnamefont {P.}~\bibnamefont
  {Wadley}}, \bibinfo {author} {\bibfnamefont {B.}~\bibnamefont {Howells}},
  \bibinfo {author} {\bibfnamefont {J.}~\bibnamefont {{\v{Z}}elezn{\'{y}}}},
  \bibinfo {author} {\bibfnamefont {C.}~\bibnamefont {Andrews}}, \bibinfo
  {author} {\bibfnamefont {V.}~\bibnamefont {Hills}}, \bibinfo {author}
  {\bibfnamefont {R.~P.}\ \bibnamefont {Campion}}, \bibinfo {author}
  {\bibfnamefont {V.}~\bibnamefont {Nov{\'{a}}k}}, \bibinfo {author}
  {\bibfnamefont {K.}~\bibnamefont {Olejn{\'{\i}}k}}, \bibinfo {author}
  {\bibfnamefont {F.}~\bibnamefont {Maccherozzi}}, \bibinfo {author}
  {\bibfnamefont {S.~S.}\ \bibnamefont {Dhesi}}, \bibinfo {author}
  {\bibfnamefont {S.~Y.}\ \bibnamefont {Martin}}, \bibinfo {author}
  {\bibfnamefont {T.}~\bibnamefont {Wagner}}, \bibinfo {author} {\bibfnamefont
  {J.}~\bibnamefont {Wunderlich}}, \bibinfo {author} {\bibfnamefont
  {F.}~\bibnamefont {Freimuth}}, \bibinfo {author} {\bibfnamefont
  {Y.}~\bibnamefont {Mokrousov}}, \bibinfo {author} {\bibfnamefont
  {J.}~\bibnamefont {Kune{\v{s}}}}, \bibinfo {author} {\bibfnamefont {J.~S.}\
  \bibnamefont {Chauhan}}, \bibinfo {author} {\bibfnamefont {M.~J.}\
  \bibnamefont {Grzybowski}}, \bibinfo {author} {\bibfnamefont {A.~W.}\
  \bibnamefont {Rushforth}}, \bibinfo {author} {\bibfnamefont {K.~W.}\
  \bibnamefont {Edmonds}}, \bibinfo {author} {\bibfnamefont {B.~L.}\
  \bibnamefont {Gallagher}}, \ and\ \bibinfo {author} {\bibfnamefont
  {T.}~\bibnamefont {Jungwirth}},\ }\href {\doibase 10.1126/science.aab1031}
  {\bibfield  {journal} {\bibinfo  {journal} {Science}\ }\textbf {\bibinfo
  {volume} {351}},\ \bibinfo {pages} {587} (\bibinfo {year}
  {2016})}\BibitemShut {NoStop}%
\bibitem [{\citenamefont {Neeraj}\ \emph {et~al.}(2020)\citenamefont {Neeraj},
  \citenamefont {Awari}, \citenamefont {Kovalev}, \citenamefont {Polley},
  \citenamefont {Hagstr\"{o}m}, \citenamefont {Arekapudi}, \citenamefont
  {Semisalova}, \citenamefont {Lenz}, \citenamefont {Green}, \citenamefont
  {Deinert}, \citenamefont {Ilyakov}, \citenamefont {Chen}, \citenamefont
  {Bawatna}, \citenamefont {Scalera}, \citenamefont {d'Aquino}, \citenamefont
  {Serpico}, \citenamefont {Hellwig}, \citenamefont {Wegrowe}, \citenamefont
  {Gensch},\ and\ \citenamefont {Bonetti}}]{Neeraj2020}%
  \BibitemOpen
  \bibfield  {author} {\bibinfo {author} {\bibfnamefont {K.}~\bibnamefont
  {Neeraj}}, \bibinfo {author} {\bibfnamefont {N.}~\bibnamefont {Awari}},
  \bibinfo {author} {\bibfnamefont {S.}~\bibnamefont {Kovalev}}, \bibinfo
  {author} {\bibfnamefont {D.}~\bibnamefont {Polley}}, \bibinfo {author}
  {\bibfnamefont {N.~Z.}\ \bibnamefont {Hagstr\"{o}m}}, \bibinfo {author}
  {\bibfnamefont {S.~S. P.~K.}\ \bibnamefont {Arekapudi}}, \bibinfo {author}
  {\bibfnamefont {A.}~\bibnamefont {Semisalova}}, \bibinfo {author}
  {\bibfnamefont {K.}~\bibnamefont {Lenz}}, \bibinfo {author} {\bibfnamefont
  {B.}~\bibnamefont {Green}}, \bibinfo {author} {\bibfnamefont {J.-C.}\
  \bibnamefont {Deinert}}, \bibinfo {author} {\bibfnamefont {I.}~\bibnamefont
  {Ilyakov}}, \bibinfo {author} {\bibfnamefont {M.}~\bibnamefont {Chen}},
  \bibinfo {author} {\bibfnamefont {M.}~\bibnamefont {Bawatna}}, \bibinfo
  {author} {\bibfnamefont {V.}~\bibnamefont {Scalera}}, \bibinfo {author}
  {\bibfnamefont {M.}~\bibnamefont {d'Aquino}}, \bibinfo {author}
  {\bibfnamefont {C.}~\bibnamefont {Serpico}}, \bibinfo {author} {\bibfnamefont
  {O.}~\bibnamefont {Hellwig}}, \bibinfo {author} {\bibfnamefont {J.-E.}\
  \bibnamefont {Wegrowe}}, \bibinfo {author} {\bibfnamefont {M.}~\bibnamefont
  {Gensch}}, \ and\ \bibinfo {author} {\bibfnamefont {S.}~\bibnamefont
  {Bonetti}},\ }\href {\doibase 10.1038/s41567-020-01040-y} {\bibfield
  {journal} {\bibinfo  {journal} {Nature Physics}\ }\textbf {\bibinfo {volume}
  {17}},\ \bibinfo {pages} {245} (\bibinfo {year} {2020})}\BibitemShut
  {NoStop}%
\bibitem [{\citenamefont {Niu}\ and\ \citenamefont
  {Kleinman}(1998{\natexlab{a}})}]{Niu1998}%
  \BibitemOpen
  \bibfield  {author} {\bibinfo {author} {\bibfnamefont {Q.}~\bibnamefont
  {Niu}}\ and\ \bibinfo {author} {\bibfnamefont {L.}~\bibnamefont {Kleinman}},\
  }\href {\doibase 10.1103/PhysRevLett.80.2205} {\bibfield  {journal} {\bibinfo
   {journal} {Phys. Rev. Lett.}\ }\textbf {\bibinfo {volume} {80}},\ \bibinfo
  {pages} {2205} (\bibinfo {year} {1998}{\natexlab{a}})}\BibitemShut {NoStop}%
\bibitem [{\citenamefont {Niu}\ \emph {et~al.}(1999{\natexlab{a}})\citenamefont
  {Niu}, \citenamefont {Wang}, \citenamefont {Kleinman}, \citenamefont {Liu},
  \citenamefont {Nicholson},\ and\ \citenamefont {Stocks}}]{Niu1999}%
  \BibitemOpen
  \bibfield  {author} {\bibinfo {author} {\bibfnamefont {Q.}~\bibnamefont
  {Niu}}, \bibinfo {author} {\bibfnamefont {X.}~\bibnamefont {Wang}}, \bibinfo
  {author} {\bibfnamefont {L.}~\bibnamefont {Kleinman}}, \bibinfo {author}
  {\bibfnamefont {W.-M.}\ \bibnamefont {Liu}}, \bibinfo {author} {\bibfnamefont
  {D.~M.~C.}\ \bibnamefont {Nicholson}}, \ and\ \bibinfo {author}
  {\bibfnamefont {G.~M.}\ \bibnamefont {Stocks}},\ }\href {\doibase
  10.1103/PhysRevLett.83.207} {\bibfield  {journal} {\bibinfo  {journal} {Phys.
  Rev. Lett.}\ }\textbf {\bibinfo {volume} {83}},\ \bibinfo {pages} {207}
  (\bibinfo {year} {1999}{\natexlab{a}})}\BibitemShut {NoStop}%
\bibitem [{\citenamefont {Zhang}\ and\ \citenamefont {Wu}(2006)}]{Zhang06}%
  \BibitemOpen
  \bibfield  {author} {\bibinfo {author} {\bibfnamefont {Q.}~\bibnamefont
  {Zhang}}\ and\ \bibinfo {author} {\bibfnamefont {B.}~\bibnamefont {Wu}},\
  }\href {\doibase 10.1103/PhysRevLett.97.190401} {\bibfield  {journal}
  {\bibinfo  {journal} {Phys. Rev. Lett.}\ }\textbf {\bibinfo {volume} {97}},\
  \bibinfo {pages} {190401} (\bibinfo {year} {2006})}\BibitemShut {NoStop}%
\bibitem [{\citenamefont {Kane}\ and\ \citenamefont {Mele}(2005)}]{Kane05}%
  \BibitemOpen
  \bibfield  {author} {\bibinfo {author} {\bibfnamefont {C.~L.}\ \bibnamefont
  {Kane}}\ and\ \bibinfo {author} {\bibfnamefont {E.~J.}\ \bibnamefont
  {Mele}},\ }\href {\doibase 10.1103/PhysRevLett.95.226801} {\bibfield
  {journal} {\bibinfo  {journal} {Phys. Rev. Lett.}\ }\textbf {\bibinfo
  {volume} {95}},\ \bibinfo {pages} {226801} (\bibinfo {year}
  {2005})}\BibitemShut {NoStop}%
\bibitem [{\citenamefont {Bernevig}\ and\ \citenamefont
  {Zhang}(2006)}]{Bernevig06}%
  \BibitemOpen
  \bibfield  {author} {\bibinfo {author} {\bibfnamefont {B.~A.}\ \bibnamefont
  {Bernevig}}\ and\ \bibinfo {author} {\bibfnamefont {S.-C.}\ \bibnamefont
  {Zhang}},\ }\href {\doibase 10.1103/PhysRevLett.96.106802} {\bibfield
  {journal} {\bibinfo  {journal} {Phys. Rev. Lett.}\ }\textbf {\bibinfo
  {volume} {96}},\ \bibinfo {pages} {106802} (\bibinfo {year}
  {2006})}\BibitemShut {NoStop}%
\bibitem [{\citenamefont {Chisnell}\ \emph {et~al.}(2015)\citenamefont
  {Chisnell}, \citenamefont {Helton}, \citenamefont {Freedman}, \citenamefont
  {Singh}, \citenamefont {Bewley}, \citenamefont {Nocera},\ and\ \citenamefont
  {Lee}}]{Chisnell15}%
  \BibitemOpen
  \bibfield  {author} {\bibinfo {author} {\bibfnamefont {R.}~\bibnamefont
  {Chisnell}}, \bibinfo {author} {\bibfnamefont {J.~S.}\ \bibnamefont
  {Helton}}, \bibinfo {author} {\bibfnamefont {D.~E.}\ \bibnamefont
  {Freedman}}, \bibinfo {author} {\bibfnamefont {D.~K.}\ \bibnamefont {Singh}},
  \bibinfo {author} {\bibfnamefont {R.~I.}\ \bibnamefont {Bewley}}, \bibinfo
  {author} {\bibfnamefont {D.~G.}\ \bibnamefont {Nocera}}, \ and\ \bibinfo
  {author} {\bibfnamefont {Y.~S.}\ \bibnamefont {Lee}},\ }\href {\doibase
  10.1103/PhysRevLett.115.147201} {\bibfield  {journal} {\bibinfo  {journal}
  {Phys. Rev. Lett.}\ }\textbf {\bibinfo {volume} {115}},\ \bibinfo {pages}
  {147201} (\bibinfo {year} {2015})}\BibitemShut {NoStop}%
\bibitem [{\citenamefont {Chen}\ \emph {et~al.}(2018)\citenamefont {Chen},
  \citenamefont {Chung}, \citenamefont {Gao}, \citenamefont {Chen},
  \citenamefont {Stone}, \citenamefont {Kolesnikov}, \citenamefont {Huang},\
  and\ \citenamefont {Dai}}]{Chen18}%
  \BibitemOpen
  \bibfield  {author} {\bibinfo {author} {\bibfnamefont {L.}~\bibnamefont
  {Chen}}, \bibinfo {author} {\bibfnamefont {J.-H.}\ \bibnamefont {Chung}},
  \bibinfo {author} {\bibfnamefont {B.}~\bibnamefont {Gao}}, \bibinfo {author}
  {\bibfnamefont {T.}~\bibnamefont {Chen}}, \bibinfo {author} {\bibfnamefont
  {M.~B.}\ \bibnamefont {Stone}}, \bibinfo {author} {\bibfnamefont {A.~I.}\
  \bibnamefont {Kolesnikov}}, \bibinfo {author} {\bibfnamefont
  {Q.}~\bibnamefont {Huang}}, \ and\ \bibinfo {author} {\bibfnamefont
  {P.}~\bibnamefont {Dai}},\ }\href {\doibase 10.1103/PhysRevX.8.041028}
  {\bibfield  {journal} {\bibinfo  {journal} {Phys. Rev. X}\ }\textbf {\bibinfo
  {volume} {8}},\ \bibinfo {pages} {041028} (\bibinfo {year}
  {2018})}\BibitemShut {NoStop}%
\bibitem [{\citenamefont {Niu}\ and\ \citenamefont
  {Kleinman}(1998{\natexlab{b}})}]{Niu98}%
  \BibitemOpen
  \bibfield  {author} {\bibinfo {author} {\bibfnamefont {Q.}~\bibnamefont
  {Niu}}\ and\ \bibinfo {author} {\bibfnamefont {L.}~\bibnamefont {Kleinman}},\
  }\href {\doibase 10.1103/PhysRevLett.80.2205} {\bibfield  {journal} {\bibinfo
   {journal} {Phys. Rev. Lett.}\ }\textbf {\bibinfo {volume} {80}},\ \bibinfo
  {pages} {2205} (\bibinfo {year} {1998}{\natexlab{b}})}\BibitemShut {NoStop}%
\bibitem [{\citenamefont {Bergqvist}\ \emph {et~al.}(2013)\citenamefont
  {Bergqvist}, \citenamefont {Taroni}, \citenamefont {Bergman}, \citenamefont
  {Etz},\ and\ \citenamefont {Eriksson}}]{Bergqvist13}%
  \BibitemOpen
  \bibfield  {author} {\bibinfo {author} {\bibfnamefont {L.}~\bibnamefont
  {Bergqvist}}, \bibinfo {author} {\bibfnamefont {A.}~\bibnamefont {Taroni}},
  \bibinfo {author} {\bibfnamefont {A.}~\bibnamefont {Bergman}}, \bibinfo
  {author} {\bibfnamefont {C.}~\bibnamefont {Etz}}, \ and\ \bibinfo {author}
  {\bibfnamefont {O.}~\bibnamefont {Eriksson}},\ }\href {\doibase
  10.1103/PhysRevB.87.144401} {\bibfield  {journal} {\bibinfo  {journal} {Phys.
  Rev. B}\ }\textbf {\bibinfo {volume} {87}},\ \bibinfo {pages} {144401}
  (\bibinfo {year} {2013})}\BibitemShut {NoStop}%
\bibitem [{\citenamefont {Savrasov}(1998)}]{Savrasov98}%
  \BibitemOpen
  \bibfield  {author} {\bibinfo {author} {\bibfnamefont {S.~Y.}\ \bibnamefont
  {Savrasov}},\ }\href {\doibase 10.1103/PhysRevLett.81.2570} {\bibfield
  {journal} {\bibinfo  {journal} {Phys. Rev. Lett.}\ }\textbf {\bibinfo
  {volume} {81}},\ \bibinfo {pages} {2570} (\bibinfo {year}
  {1998})}\BibitemShut {NoStop}%
\bibitem [{\citenamefont {Rousseau}\ \emph {et~al.}(2012)\citenamefont
  {Rousseau}, \citenamefont {Eiguren},\ and\ \citenamefont
  {Bergara}}]{Rousseau12}%
  \BibitemOpen
  \bibfield  {author} {\bibinfo {author} {\bibfnamefont {B.}~\bibnamefont
  {Rousseau}}, \bibinfo {author} {\bibfnamefont {A.}~\bibnamefont {Eiguren}}, \
  and\ \bibinfo {author} {\bibfnamefont {A.}~\bibnamefont {Bergara}},\ }\href
  {\doibase 10.1103/PhysRevB.85.054305} {\bibfield  {journal} {\bibinfo
  {journal} {Phys. Rev. B}\ }\textbf {\bibinfo {volume} {85}},\ \bibinfo
  {pages} {054305} (\bibinfo {year} {2012})}\BibitemShut {NoStop}%
\bibitem [{\citenamefont {Singh}\ \emph {et~al.}(2019)\citenamefont {Singh},
  \citenamefont {Elliott}, \citenamefont {Nautiyal}, \citenamefont {Dewhurst},\
  and\ \citenamefont {Sharma}}]{Singh19}%
  \BibitemOpen
  \bibfield  {author} {\bibinfo {author} {\bibfnamefont {N.}~\bibnamefont
  {Singh}}, \bibinfo {author} {\bibfnamefont {P.}~\bibnamefont {Elliott}},
  \bibinfo {author} {\bibfnamefont {T.}~\bibnamefont {Nautiyal}}, \bibinfo
  {author} {\bibfnamefont {J.~K.}\ \bibnamefont {Dewhurst}}, \ and\ \bibinfo
  {author} {\bibfnamefont {S.}~\bibnamefont {Sharma}},\ }\href {\doibase
  10.1103/PhysRevB.99.035151} {\bibfield  {journal} {\bibinfo  {journal} {Phys.
  Rev. B}\ }\textbf {\bibinfo {volume} {99}},\ \bibinfo {pages} {035151}
  (\bibinfo {year} {2019})}\BibitemShut {NoStop}%
\bibitem [{\citenamefont {Tancogne-Dejean}\ \emph {et~al.}(2020)\citenamefont
  {Tancogne-Dejean}, \citenamefont {Eich},\ and\ \citenamefont
  {Rubio}}]{TancogneDejean20}%
  \BibitemOpen
  \bibfield  {author} {\bibinfo {author} {\bibfnamefont {N.}~\bibnamefont
  {Tancogne-Dejean}}, \bibinfo {author} {\bibfnamefont {F.~G.}\ \bibnamefont
  {Eich}}, \ and\ \bibinfo {author} {\bibfnamefont {A.}~\bibnamefont {Rubio}},\
  }\href {\doibase 10.1021/acs.jctc.9b01064} {\bibfield  {journal} {\bibinfo
  {journal} {Journal of Chemical Theory and Computation}\ } (\bibinfo {year}
  {2020}),\ 10.1021/acs.jctc.9b01064}\BibitemShut {NoStop}%
\bibitem [{\citenamefont {Lieb}(1973)}]{Lieb73}%
  \BibitemOpen
  \bibfield  {author} {\bibinfo {author} {\bibfnamefont {E.~H.}\ \bibnamefont
  {Lieb}},\ }\href {\doibase 10.1007/bf01646493} {\bibfield  {journal}
  {\bibinfo  {journal} {Communications in Mathematical Physics}\ }\textbf
  {\bibinfo {volume} {31}},\ \bibinfo {pages} {327} (\bibinfo {year}
  {1973})}\BibitemShut {NoStop}%
\bibitem [{\citenamefont {Fradkin}(2013)}]{Fradkin13}%
  \BibitemOpen
  \bibfield  {author} {\bibinfo {author} {\bibfnamefont {E.}~\bibnamefont
  {Fradkin}},\ }\href {\doibase 10.1017/CBO9781139015509} {\emph {\bibinfo
  {title} {Field Theories of Condensed Matter Physics}}},\ \bibinfo {edition}
  {2nd}\ ed.\ (\bibinfo  {publisher} {Cambridge University Press},\ \bibinfo
  {year} {2013})\BibitemShut {NoStop}%
\bibitem [{\citenamefont {Stahl}\ and\ \citenamefont
  {Potthoff}(2017)}]{Stahl17}%
  \BibitemOpen
  \bibfield  {author} {\bibinfo {author} {\bibfnamefont {C.}~\bibnamefont
  {Stahl}}\ and\ \bibinfo {author} {\bibfnamefont {M.}~\bibnamefont
  {Potthoff}},\ }\href {\doibase 10.1103/PhysRevLett.119.227203} {\bibfield
  {journal} {\bibinfo  {journal} {Phys. Rev. Lett.}\ }\textbf {\bibinfo
  {volume} {119}},\ \bibinfo {pages} {227203} (\bibinfo {year}
  {2017})}\BibitemShut {NoStop}%
\bibitem [{\citenamefont {Elbracht}\ \emph {et~al.}(2020)\citenamefont
  {Elbracht}, \citenamefont {Michel},\ and\ \citenamefont
  {Potthoff}}]{Elbracht2020}%
  \BibitemOpen
  \bibfield  {author} {\bibinfo {author} {\bibfnamefont {M.}~\bibnamefont
  {Elbracht}}, \bibinfo {author} {\bibfnamefont {S.}~\bibnamefont {Michel}}, \
  and\ \bibinfo {author} {\bibfnamefont {M.}~\bibnamefont {Potthoff}},\ }\href
  {\doibase 10.1103/PhysRevLett.124.197202} {\bibfield  {journal} {\bibinfo
  {journal} {Phys. Rev. Lett.}\ }\textbf {\bibinfo {volume} {124}},\ \bibinfo
  {pages} {197202} (\bibinfo {year} {2020})}\BibitemShut {NoStop}%
\bibitem [{\citenamefont {Altland}\ and\ \citenamefont
  {Simons}(2010)}]{Altland10}%
  \BibitemOpen
  \bibfield  {author} {\bibinfo {author} {\bibfnamefont {A.}~\bibnamefont
  {Altland}}\ and\ \bibinfo {author} {\bibfnamefont {B.~D.}\ \bibnamefont
  {Simons}},\ }\href {\doibase 10.1017/CBO9780511789984} {\emph {\bibinfo
  {title} {Condensed Matter Field Theory}}},\ \bibinfo {edition} {2nd}\ ed.\
  (\bibinfo  {publisher} {Cambridge University Press},\ \bibinfo {year}
  {2010})\BibitemShut {NoStop}%
\bibitem [{\citenamefont {Qian}\ and\ \citenamefont {Vignale}(2002)}]{Qian02}%
  \BibitemOpen
  \bibfield  {author} {\bibinfo {author} {\bibfnamefont {Z.}~\bibnamefont
  {Qian}}\ and\ \bibinfo {author} {\bibfnamefont {G.}~\bibnamefont {Vignale}},\
  }\href {\doibase 10.1103/PhysRevLett.88.056404} {\bibfield  {journal}
  {\bibinfo  {journal} {Phys. Rev. Lett.}\ }\textbf {\bibinfo {volume} {88}},\
  \bibinfo {pages} {056404} (\bibinfo {year} {2002})}\BibitemShut {NoStop}%
\bibitem [{\citenamefont {Niu}\ \emph {et~al.}(1999{\natexlab{b}})\citenamefont
  {Niu}, \citenamefont {Wang}, \citenamefont {Kleinman}, \citenamefont {Liu},
  \citenamefont {Nicholson},\ and\ \citenamefont {Stocks}}]{Niu99}%
  \BibitemOpen
  \bibfield  {author} {\bibinfo {author} {\bibfnamefont {Q.}~\bibnamefont
  {Niu}}, \bibinfo {author} {\bibfnamefont {X.}~\bibnamefont {Wang}}, \bibinfo
  {author} {\bibfnamefont {L.}~\bibnamefont {Kleinman}}, \bibinfo {author}
  {\bibfnamefont {W.-M.}\ \bibnamefont {Liu}}, \bibinfo {author} {\bibfnamefont
  {D.~M.~C.}\ \bibnamefont {Nicholson}}, \ and\ \bibinfo {author}
  {\bibfnamefont {G.~M.}\ \bibnamefont {Stocks}},\ }\href {\doibase
  10.1103/PhysRevLett.83.207} {\bibfield  {journal} {\bibinfo  {journal} {Phys.
  Rev. Lett.}\ }\textbf {\bibinfo {volume} {83}},\ \bibinfo {pages} {207}
  (\bibinfo {year} {1999}{\natexlab{b}})}\BibitemShut {NoStop}%
\bibitem [{\citenamefont {Overhauser}(1962)}]{Overhauser62}%
  \BibitemOpen
  \bibfield  {author} {\bibinfo {author} {\bibfnamefont {A.~W.}\ \bibnamefont
  {Overhauser}},\ }\href {\doibase 10.1103/PhysRev.128.1437} {\bibfield
  {journal} {\bibinfo  {journal} {Phys. Rev.}\ }\textbf {\bibinfo {volume}
  {128}},\ \bibinfo {pages} {1437} (\bibinfo {year} {1962})}\BibitemShut
  {NoStop}%
\bibitem [{SM()}]{SM}%
  \BibitemOpen
  \href@noop {} {}\bibinfo {note} {See Supplemental Material at [URL] for a
  detailed derivation of the electronic spin Berry curvature.}\BibitemShut
  {Stop}%
\bibitem [{\citenamefont {Eich}\ \emph {et~al.}(2013)\citenamefont {Eich},
  \citenamefont {Pittalis},\ and\ \citenamefont {Vignale}}]{Eich13}%
  \BibitemOpen
  \bibfield  {author} {\bibinfo {author} {\bibfnamefont {F.~G.}\ \bibnamefont
  {Eich}}, \bibinfo {author} {\bibfnamefont {S.}~\bibnamefont {Pittalis}}, \
  and\ \bibinfo {author} {\bibfnamefont {G.}~\bibnamefont {Vignale}},\ }\href
  {\doibase 10.1103/PhysRevB.88.245102} {\bibfield  {journal} {\bibinfo
  {journal} {Phys. Rev. B}\ }\textbf {\bibinfo {volume} {88}},\ \bibinfo
  {pages} {245102} (\bibinfo {year} {2013})}\BibitemShut {NoStop}%
\bibitem [{\citenamefont {Baer}(2006)}]{Baer06}%
  \BibitemOpen
  \bibfield  {author} {\bibinfo {author} {\bibfnamefont {M.}~\bibnamefont
  {Baer}},\ }\href {\doibase 10.1002/0471780081} {\emph {\bibinfo {title}
  {Beyond Born-Oppenheimer}}}\ (\bibinfo  {publisher} {John Wiley {\&} Sons,
  Inc.},\ \bibinfo {year} {2006})\BibitemShut {NoStop}%
\bibitem [{\citenamefont {Park}\ and\ \citenamefont {Light}(1986)}]{Park86}%
  \BibitemOpen
  \bibfield  {author} {\bibinfo {author} {\bibfnamefont {T.~J.}\ \bibnamefont
  {Park}}\ and\ \bibinfo {author} {\bibfnamefont {J.~C.}\ \bibnamefont
  {Light}},\ }\href {\doibase 10.1063/1.451548} {\bibfield  {journal} {\bibinfo
   {journal} {The Journal of Chemical Physics}\ }\textbf {\bibinfo {volume}
  {85}},\ \bibinfo {pages} {5870} (\bibinfo {year} {1986})},\ \Eprint
  {http://arxiv.org/abs/https://doi.org/10.1063/1.451548}
  {https://doi.org/10.1063/1.451548} \BibitemShut {NoStop}%
\bibitem [{\citenamefont {Depondt}\ and\ \citenamefont
  {Mertens}(2009)}]{Depondt09}%
  \BibitemOpen
  \bibfield  {author} {\bibinfo {author} {\bibfnamefont {P.}~\bibnamefont
  {Depondt}}\ and\ \bibinfo {author} {\bibfnamefont {F.~G.}\ \bibnamefont
  {Mertens}},\ }\href {\doibase 10.1088/0953-8984/21/33/336005} {\bibfield
  {journal} {\bibinfo  {journal} {Journal of Physics: Condensed Matter}\
  }\textbf {\bibinfo {volume} {21}},\ \bibinfo {pages} {336005} (\bibinfo
  {year} {2009})}\BibitemShut {NoStop}%
\bibitem [{Nic()}]{Nicolas}%
  \BibitemOpen
  \href@noop {} {}\bibinfo {note} {Nicolas Tancogne-Dejean and Angel Rubio
  (private communications)}\BibitemShut {NoStop}%
\end{thebibliography}%

\end{document}